\documentclass[12pt,notitlepage]{article}
\usepackage{amsmath,amssymb,amsthm}
\usepackage{cite}
\usepackage{graphics}
\usepackage{latexsym}
\usepackage{amsmath}
\usepackage{amscd}
\usepackage{graphicx}
\begin{document}
 \vspace{15pt}

\begin{center}{\Large \bf The new integrable deformations of short pulse equation and sine-Gordon equation, and their solutions}
\end{center}
\begin{center}
{\it Yuqin Yao$^{1)}$\footnote{Corresponding author: yyqinw@126.com
}, Yehui Huang$^{2)}$\footnote{huangyh@mails.tsinghua.edu.cn},
Guixiang Dong$^{3)}$ and Yunbo
Zeng$^{2)}$\footnote{yzeng@math.tsinghua.edu.cn} }
\end{center}
\begin{center}{\small \it$^{1)}$Department of
  Applied Mathematics, China Agricultural University, Beijing, 100083, PR China\\
 $^{2)}$Department of Mathematical Science,
Tsinghua University, Beijing, 100084 , PR China\\
$^{3)}$ College of Science,Shandong Jianzhu University, Shandong
Jinan 250101 PR China\\}
\end{center}

\vskip 12pt { \small\noindent\bf Abstract}
 {We first derive an integrable deformed hierarchy of short pulse
 equation and their Lax representation. Then we concentrated on the
 solution of integrable deformed short pulse equation (IDSPE). By
 proposing a generalized reciprocal transformation, we find a new integrable
 deformed sine-Gordon equation (IDSGE) and its Lax representation.
 The multisoliton
  solutions, negaton solutions and positon solutions for the IDSGE and the
  N-loop soliton
solutions, N- negaton and N-positon solutions for
  the IDSPE are presented. In the reduced case the new N-positon solutions and N-negaton solutions for
  short pulse equation are obtained.}

\section{Introduction}
It is known that pulse propagation in optical fibers is usually
modeled by the cubic nonlinear Schr$\ddot{o}$dinger
equaion\cite{ns}. However, it is not valid for the ultra-short
pules. In 2004, starting from the Maxwell equation of electric field
in the fiber, Sch$\ddot{a}$fer and Wayne derived the short pulse
equation(SPE)\cite{sp}
\begin{equation}
\label{i1} u_{xt}=u+\frac{1}{6}(u^{3})_{xx}
\end{equation}
as an alternative of the cubic nonlinear Schr$\ddot{o}$dinger
equation to describing the propagation of ultrashort optical pules
in nonlinear media, where  $u(x,t)$ represents the magnitude of
electric field. In recent years, the short pulse model has attracted
considerable attention. In \cite{ad1,ad2}, it was also deduced from
the partial differential equation which describes pseudospherical
surfaces. In \cite{laxp}, Sakovich and Sakovich proved that the SPE
is integrable by discovering a Lax pair for the SPE. Brunelli proved
the integrability of the SPE from the Hamiltonian point of view and
studied the short pulse hierarchy in \cite{hm1,hm2}. In \cite{disc},
Feng et al. proposed the integrable semi-discrete and full-discrete
analogues of the SPE. As far as the solutions of the SPE are
considered, in \cite{solu1} not only solitary wave solutions are
obtained by making use of transformation between the SPE and
sine-Gordon equation but also the pulse solutions of the SPE were
derived from the breather solutions of the SG equation. Some
periodic and traveling wave solutions of SPE are given in
\cite{solu2}. Kuetche et al. constructed the two loop soliton
solutions with use of bilinear method and hodograph
transformation\cite{solu3}. In\cite{solu4,solu5}, Matsuno develop a
systematic procedure to construct the periodic solutions and
multiloop solitons.

The integrable deformation of integrable system attract a lot of
interests from both physical and mathematical points. One kind of
the integrable deformation is the so-called soliton equation with
self-consistent sources\cite{a1}-\cite{scs10}, which consists of the
soliton equation with additional terms by coupling the corresponding
eigenvalue problems, and has important application. For example, the
nonlinear Schr$\ddot{o}$dinger equation with self-consistent sources
is relevant to some problems of plasma physical and solid state
physics.

In this paper, we consider the integrable deformed short pulse
equation (IDSPE) which has not been studied yet. First we derive the
integrable deformed hierarchy of short pulse equation(IDSPH) and
their Lax representation, which includes integrable deformed short
pulse equation (IDSPE) and its Lax representation. This implies that
the IDSPE is Lax integrable.
 A generalized reciprocal transformation for the IDSPE is proposed. This
 transformation
converts the IDSPE  and its Lax representation into new integrable
deformed sine-Gordon equation (IDSGE) and its Lax representation.
The IDSGE can be written as bilinear form by introducing an
independent variable transformation. The N-soliton solutions of the
sine-Gordon equation were obtained in \cite{hirota, beu}, and the
positon solutions of sine-Gordon equation and its properties were
studied in \cite{beu}. Here, we find the N-soliton solutions,
N-negatons solutions and N-positon solutions for the IDSGE .
 Further using the inverse
reciprocal transformation, we construct N-loop soliton solutions,
N-negaton solutions and N-positon solutions of the IDSPE. In the
reduced case we obtain the new N-negaton solutions and new N-positon
solutions for the short pulse equation.

This paper is organized as follows. In section 2, we establish the
IDSPH and its Lax representation. In section 3,  A generalized
reciprocal transformation for the IDSPE is proposed and the IDSGE is
worked out.  In section 4, the  solutions of the IDSGE are obtained.
Section 5 gives the solutions  of IDSPE. The conclusion is given in
section 6.

\section{The new integrable deformed hierarchy of  short pulse equation (IDSPH) and  its Lax pair}
\subsection{The IDSPH}
Consider the eigenvalue problem\cite{laxp}
\begin{equation}
\label{sp1} \left(\begin{array}{c}
\varphi_{1}\\
\varphi_{2}\\
 \end{array}\right)_{x}=U\left(\begin{array}{c}
\varphi_{1}\\
\varphi_{2}\\
 \end{array}\right),~~U=\left(\begin{array}{cc}
\lambda & \lambda u_x\\
\lambda u_x & -\lambda\\
 \end{array}\right).
\end{equation}
The adjoint representation reads
\begin{equation}
\label{sp2} V_x=[U,V].
\end{equation}
 Set
\begin{equation}
\label{sp4}V=\left(\begin{array}{cc}
A & B\\
C & -A\\
 \end{array}\right)=\sum_{m=0}^{\infty}\left(\begin{array}{cc}
\lambda a_m & \lambda u_x a_m+b_m\\
\lambda u_x a_m+c_m & -\lambda a_m\\
 \end{array}\right)\lambda^{m}.
\end{equation}
Eq. (\ref{sp2}) yields
\begin{equation}
\label{sp5}  \begin{cases} a_{m,x}=u_x(c_m-b_m),\\
 b_{m+1,x}=2b_m-(u_xa_m)_x,\\
c_{m+1,x}=-2c_m-(u_xa_m)_x.
  \end{cases}
\end{equation}
Taking $b_0=c_0=0,~a_0=\frac{1}{4}$, we have
\begin{equation}
\label{spad1}
\begin{cases}
~b_1=c_1=-\frac{1}{4}u_x,~a_1=0,~\\
b_2=-c_2=-\frac{1}{2}u,~a_2=\frac{1}{2}u^{2},\cdots,
 \end{cases}
\end{equation}
and  in general,
\begin{subequations}
\label{sp6}
\begin{align}
& b_{2n}=-c_{2n}=2\partial^{-1}b_{2n-1},\\
&a_{2n}=-2\partial^{-1}(u_xb_{2n}),~a_{2n+1}=0,\\
&b_{2n+1}=c_{2n+1}=2\partial^{-1}b_{2n}-u_xa_{2n}=Lb_{2n-1},~L=4(\partial^{-1}+u_x\partial^{-1}u_x)\partial^{-1},
\end{align}
\end{subequations}
where $\partial=\frac{\partial}{\partial
x},~\partial\partial^{-1}=\partial^{-1}\partial=1$.\\
Set \begin{equation} \label{spad2}
V^{(n)}=\sum_{m=0}^{2n}\left(\begin{array}{cc}
\lambda a_m & \lambda u_x a_m+b_m\\
\lambda u_x a_m+c_m & -\lambda a_m\\
 \end{array}\right)\lambda^{m-2n},
\end{equation}
 and take
\begin{equation}
\label{sp8}\left(\begin{array}{c}
\varphi_{1}\\
\varphi_{2}\\
 \end{array}\right)_{t_{n}}=V^{(n)}\left(\begin{array}{c}
\varphi_{1}\\
\varphi_{2}\\
\end{array}\right).\end{equation}
Then the compatibility condition of Eqs.(\ref{sp1}) and (\ref{sp8})
gives rise to the short pulse hierarchy(SPH)
\begin{equation}
\label{sp9}u_{xt_n}=-\partial b_{2n+1},
~n=0,1,~\cdots.\end{equation} When $n=1$, (\ref{sp9}) gives the
short pulse equation
\begin{equation}
\label{sp10}u_{xt}=u+\frac{1}{6}(u^{3})_{xx},\end{equation} and
$V^{(2)}$ in (\ref{sp8}) is given by
 \begin{equation}
 \label{spad3}
 V^{(2)}= \left(\begin{array}{cc}
\frac{1}{2}\lambda u^{2}+\frac{1}{4\lambda} & \frac{1}{2}\lambda u^{2}u_x-\frac{u}{2} \\
\frac{1}{2}\lambda u^{2}u_x+\frac{u}{2}& -\frac{1}{2}\lambda u^{2}-\frac{1}{4\lambda} \\
 \end{array}\right).
\end{equation}
For $n$ distinct real $\lambda_{j}$, consider the following spectral
problem
$$
  \left(\begin{array}{c}
\varphi_{1j}\\
\varphi_{2j}\\
 \end{array}\right)_{x}=\left(\begin{array}{cc}
\lambda_j & \lambda_j u_x\\
\lambda_j u_x & -\lambda_j\\
 \end{array}\right)\left(\begin{array}{c}
\varphi_{1j}\\
\varphi_{2j}\\
 \end{array}\right).
$$
It is easy to find that
\begin{equation}
 \label{sp12}\frac{\delta \lambda_j}{\delta
u}=-2\lambda_j(\varphi_{1j}^{2}+\varphi_{2j}^{2}),~L(\varphi_{1j}^{2}+\varphi_{2j}^{2})_x=\frac{1}{\lambda_j^{2}}(\varphi_{1j}^{2}
+\varphi_{2j}^{2})_x.
\end{equation}
According to the approach proposed in Refs.\cite{a4}-\cite{scs10},
the short pulse hierarchy with self-consistent sources (SPHSCS) or
integrable deformed short pulse hierarchy (IDSPH) is defined by
\begin{subequations}
\label{sp13}
\begin{align}
& u_{xt_n}=-\partial[b_{2n+1}-\sum_{j=1}^{N}\frac{1}{
2\lambda_j^{2}}(\varphi_{1j}^{2}+\varphi_{2j}^{2})_x],\\
&\varphi_{1jx}=\lambda_j\varphi_{1j}+\lambda_j u_x
\varphi_{2j},~\varphi_{2jx}=\lambda_j u_x \varphi_{1j}-\lambda_j
\varphi_{2j},~j=1,2,\cdots,N.
\end{align}
\end{subequations}
When $n=1$, (\ref{sp13})  gives the short pulse equation with
self-consistent sources (SPESCS) or integrable deformed short pulse
equation (IDSPE)
\begin{subequations}
\label{sp14}
\begin{align}
& u_{xt}=u+\frac{1}{6}(u^{3})_{xx}+\sum_{j=1}^{N}\frac{1}{
2\lambda_j^{2}}(\varphi_{1j}^{2}+\varphi_{2j}^{2})_{xx}, \label{sp14aa}\\
&\varphi_{1jx}=\lambda_j\varphi_{1j}+\lambda_j u_x
\varphi_{2j},~\varphi_{2jx}=\lambda_j u_x \varphi_{1j}-\lambda_j
\varphi_{2j},~j=1,2,\cdots,N. \label{sp14bb}
\end{align}
\end{subequations}
\subsection{ Lax pair of the IDSPH }
In order to find the Lax pair for IDSPE (\ref{sp14}), we first
consider the following stationary equation of  (\ref{sp14}).
\begin{subequations}
\label{spadad}
\begin{align}
&b_{3}-\sum_{j=1}^{N}\frac{1}{
2\lambda_j^{2}}(\varphi_{1j}^{2}+\varphi_{2j}^{2})_x=0,\label{spadada}\\
&\varphi_{1jx}=\lambda_j\varphi_{1j}+\lambda_j u_x
\varphi_{2j},~\varphi_{2jx}=\lambda_j u_x \varphi_{1j}-\lambda_j
\varphi_{2j},~j=1,2,\cdots,N.\label{spadadb}
\end{align}
\end{subequations}
According to Eqs.(\ref{spad1}), ~(\ref{sp6}),~(\ref{sp12}) and
(\ref{spadad}), we may define
$$\bar{a}_0=\frac{1}{4},~\bar{b}_0=\bar{c}_0=0,~\bar{b}_1=\bar{c}_1=-\frac{1}{4}u_x,~\bar{a}_1=0,~
\bar{b}_2=-\bar{c}_2=-\frac{1}{2}u,~\bar{a}_2=\frac{1}{2}u^{2},~~~~~~~~~~~~~~~~$$
$$\bar{b}_{2n+1}=\bar{c}_{2n+1}=L^{n-1}\bar{b}_3=L^{n-1}\sum_{j=1}^{N}\frac{1}{
2\lambda_j^{2}}(\varphi_{1j}^{2}+\varphi_{2j}^{2})_x=
\sum_{j=1}^{N}\frac{1}{
2\lambda_j^{2n}}(\varphi_{1j}^{2}+\varphi_{2j}^{2})_x,~n=1,2,\cdots,
$$$$\bar{b}_{2n}=-\bar{c}_{2n}=2\partial^{-1}\bar{b}_{2n-1}=\sum_{j=1}^{N}\frac{1}{
\lambda_j^{2n-2}}(\varphi_{1j}^{2}+\varphi_{2j}^{2}),~n=2,3,\cdots,~~~~~~~~~~~~~~~~~~~~~~~~~~~~~~~~~~$$

$$\bar{a}_{2n}=-2\partial^{-1}(u_x\bar{b}_{2n})=-2\partial^{-1}\sum_{j=1}^{N}\frac{1}{
\lambda_j^{2n-2}}u_x(\varphi_{1j}^{2}+\varphi_{2j}^{2})=-2\sum_{j=1}^{N}\frac{1}{
\lambda_j^{2n-1}}\varphi_{1j}\varphi_{2j},~~n=2,3,\cdots,$$
\begin{equation}
 \label{spad4}
\bar{a}_{2n+1}=0,~~n=1,2,3,\cdots.~~~~~~~~~~~~~~~~~~~~~~~~~~~~~~~~~~~~~~~~~~~~~~~~~~~~~~~~~~~~~~~~~~~~~~~~~~~~~
\end{equation}
Then  we have
$$\bar{A}=\lambda^{-2}\sum_{n=0}^{\infty}\bar{a}_{n}\lambda^{n+1}=
\frac{1}{4\lambda} + \frac{1}{2}u^{2}+\bar{A}_0,$$
$$\bar{A}_0=\sum_{n=2}^{\infty}\bar{a}_{2n}\lambda^{2n-1}=-2\sum_{j=1}^{N}\sum_{n=2}^{\infty}(\frac{\lambda}{
\lambda_j})^{2n-1}\varphi_{1j}\varphi_{2j}=2\lambda\sum_{j=1}^{N}\frac{1}{\lambda_j}\varphi_{1j}\varphi_{2j}+
2\lambda\sum_{j=1}^{N}\frac{\lambda_j}{\lambda^{2}-\lambda_j^{2}}\varphi_{1j}\varphi_{2j}.$$
In the same way, we find that
$$\bar{V}=\left(\begin{array}{cc}
\bar{A} & \bar{B} \\
\bar{C} & -\bar{A}\\
 \end{array}\right)=\lambda^{-2}\sum_{n=0}^{\infty}\left(\begin{array}{cc}
\lambda \bar{a}_{n}  & \lambda u_x \bar{a}_{n}+\bar{b}_{n} \\
 \lambda u_x \bar{a}_{n}+\bar{c}_{n} & -\lambda \bar{a}_{n}\\
 \end{array}\right)\lambda^{n}$$
$$=V^{(2)}+N_0,~~~N_0=\left(\begin{array}{cc}
\bar{A}_0 & \bar{B}_0 \\
\bar{C}_0 & -\bar{A}_0\\
 \end{array}\right).~~~~~~~~~~~~~~~~~~~~~~~$$
$$
\bar{B}_0=\sum_{j=1}^{N}[\frac{2\lambda}{\lambda_j}u_x\varphi_{1j}\varphi_{2j}-(\varphi_{1j}^{2}+\varphi_{2j}^{2})
-\frac{\lambda_j}{\lambda}(\varphi_{1j}^{2}-\varphi_{2j}^{2})+\frac{\lambda_j^{2}
}{\lambda^{2}-\lambda_j^{2}}[\frac{\lambda_j}{\lambda}(\varphi_{2j}^{2}-\varphi_{1j}^{2})-(\varphi_{1j}^{2}+\varphi_{2j}^{2})]].$$
$$\bar{C}_0=\sum_{j=1}^{N}[\frac{2\lambda}{\lambda_j}u_x\varphi_{1j}\varphi_{2j}+(\varphi_{1j}^{2}+\varphi_{2j}^{2})
-\frac{\lambda_j}{\lambda}(\varphi_{1j}^{2}-\varphi_{2j}^{2})+\frac{\lambda_j^{2}
}{\lambda^{2}-\lambda_j^{2}}[\frac{\lambda_j}{\lambda}(\varphi_{2j}^{2}-\varphi_{1j}^{2})+(\varphi_{1j}^{2}+\varphi_{2j}^{2})]].$$
Since $\bar{a}_{n},~\bar{b}_{n}$ and $\bar{c}_{n}$ satisfy the same
recursion relations as (\ref{sp5}). It is obvious that $\bar{V}$
satisfies
\begin{equation}
 \label{spad5}
\bar{V}_x=[U,\bar{V}].
\end{equation}
In fact, it is easy to verify that  (\ref{spad5}) under
(\ref{spadadb}) leads to  (\ref{spadada}). Since (\ref{spadad}) is
the stationary equation of (\ref{sp14}), we immediately obtain the
zero curvature representation for IDSPE (\ref{sp14})
\begin{equation}
 \label{spad6}
U_t-\bar{V}_x+[U,\bar{V}]=0.
\end{equation}
with the Lax pair for the IDSPE (\ref{sp14})
\begin{subequations}
\label{sp15}
\begin{align}
& \left(\begin{array}{c}
\phi_{1}\\
\phi_{2}\\
 \end{array}\right)_{x}=\left(\begin{array}{cc}
\lambda & \lambda u_x\\
\lambda u_x & -\lambda\\
 \end{array}\right)\left(\begin{array}{c}
\phi_{1}\\
\phi_{2}\\
 \end{array}\right),\\
&\left(\begin{array}{c}
\phi_{1}\\
\phi_{2}\\
 \end{array}\right)_{t}=\bar{V}\left(\begin{array}{c}
\phi_{1}\\
\phi_{2}\\
 \end{array}\right).
\end{align}
\end{subequations}
Furthermore, the zero curvature representation  and Lax pair for
IDSPH (\ref{sp13}) are given by (\ref{spad6}) and  (\ref{sp15}) with
\begin{equation}
 \label{spad7}
 \bar{V}=V^{(n)}+N_0.\end{equation}
{\bf Remark~1.} The zero curvature representation and Lax pair for
IDSPH (\ref{sp13}) are given by  (\ref{spad6}),~(\ref{sp15}) and
(\ref{spad7}). This implies that the new IDSPH is integrable in the
Lax sense.

\section{The new integrable deformed sine-Gordon equation (IDSGE) }
By introducing the new dependent variable \cite{solu4}
\begin{equation}
\label{sg1} r^{2}=1+u_x^{2}.\end{equation} Eq.(\ref{sp14aa}) is
transformed into the form
\begin{equation}
\label{sg2}
r_t=\frac{1}{2}(u^{2}r)_x+\frac{u_x}{2r}\sum_{j=1}^{N}\lambda_j^{-2}(\varphi_{1j}^{2}+\varphi_{2j}^{2})_{xx}
=(\frac{1}{2}u^{2}r+2r\sum_{j=1}^{N}\lambda_j^{-1}\varphi_{1j}\varphi_{2j})_x.
\end{equation}
 So we can define a reciprocal transformation
$(x,t)\rightarrow (y,s)$ by the relation
\begin{equation}
\label{sg3} dy=r
dx+(\frac{1}{2}u^{2}r+2r\sum_{j=1}^{N}\lambda_j^{-1}\varphi_{1j}\varphi_{2j})ds,~~~ds=dt,
\end{equation}
and we have
\begin{equation}
\label{sg4} \frac{\partial}{\partial x}=r\frac{\partial}{\partial
y},~~~~\frac{\partial}{\partial t}=\frac{\partial}{\partial
s}+(\frac{1}{2}u^{2}r+2r\sum_{j=1}^{N}\lambda_j^{-1}\varphi_{1j}\varphi_{2j})\frac{\partial}{\partial
y}.
\end{equation}
Denoting
$\phi_i(x,t)=\psi_i(y,s),~\varphi_{ij}(x,t)=\psi_{ij}(y,s),~(i=1,2)$,
with the new variable $y$ and $s$, (\ref{sg1}) and (\ref{sg2}) are
transformed into
\begin{equation}
\label{sg5} r^{2}=1+r^{2}u_y^{2},
\end{equation}
\begin{equation}
\label{sg6}
r_s=r^{2}uu_y+2r^{2}\sum_{j=1}^{N}\lambda_j^{-1}(\psi_{1j}\psi_{2j})_y,
\end{equation}
respectively. Furthermore, we define
\begin{equation}
\label{sg7} u_y=sinz, ~z=z(y,s).
\end{equation}
Inserting (\ref{sg7}) into (\ref{sg5}) gives rise to
\begin{equation}
\label{sg8}r=\frac{1}{cosz}.
\end{equation}
Using (\ref{sg4}) and (\ref{sg8}),  (\ref{sp14bb}) is converted the
following form
\begin{equation}
\label{sg9} \psi_{1jy}=\lambda_jcosz \psi_{1j}+\lambda_j
sinz\psi_{2j},~\psi_{2jy}=\lambda_j sinz\psi_{1j}-\lambda_j
cosz\psi_{2j},~j=1,2,\cdots,N.
\end{equation}
Under Eqs.(\ref{sg7})-(\ref{sg9}), (\ref{sg6}) becomes
\begin{equation}
\label{sgad1}z_s=u+2\sum_{j=1}^{N}(\psi_{1j}^{2}+\psi_{2j}^{2}).\end{equation}
So under the reciprocal transformation (\ref{sg4}), the IDSPE
(\ref{sp14}) is transformed into
 the following  integrable deformed sine-Gordon
equation (IDSGE)
\begin{subequations}
\label{sg10}
\begin{align}
& z_{ys}=sinz+2\sum_{j=1}^{N}(\psi_{1j}^{2}+\psi_{2j}^{2})_y, \label{sg10a}\\
&\psi_{1jy}=\lambda_jcosz \psi_{1j}+\lambda_j
sinz\psi_{2j},~\psi_{2jy}=\lambda_j sinz\psi_{1j}-\lambda_j
cosz\psi_{2j},~j=1,2,\cdots,N. \label{sg10b}
\end{align}
\end{subequations}
Under the reciprocal transformation (\ref{sg4}), (\ref{sg7}),
(\ref{sg8}) and (\ref{sgad1}), the Lax pair (\ref{sp15}) for the
IDSPE (\ref{sp14}) are transformed into the Lax pair for
(\ref{sg10})
\begin{subequations}
\label{sg11}
\begin{align}
& \left(\begin{array}{c}
\psi_{1}\\
\psi_{2}\\
 \end{array}\right)_{y}=\left(\begin{array}{cc}
\lambda cosz& \lambda sinz\\
\lambda sinz & -\lambda cosz\\
 \end{array}\right)\left(\begin{array}{c}
\psi_{1}\\
\psi_{2}\\
 \end{array}\right),\\
&\left(\begin{array}{c}
\psi_{1}\\
\psi_{2}\\
 \end{array}\right)_{s}=N\left(\begin{array}{c}
\psi_{1}\\
\psi_{2}\\
 \end{array}\right),
\end{align}
\end{subequations}
$$N=\left(\begin{array}{cc}
\frac{1}{4\lambda} &-\frac{1}{2}z_s\\
\frac{1}{2}z_s &-\frac{1}{4\lambda}\\
 \end{array}\right)~~~~~~~~~~~~~~~~~~~
 $$$$+\sum_{j=1}^{N}\frac{\lambda_j}{\lambda^{2}-\lambda_j^{2}}\left(\begin{array}{cc}
2\lambda \psi_{1j}\psi_{2j} &-\lambda_j(\psi_{1j}^{2}+\psi_{2j}^{2})-\lambda(\psi_{1j}^{2}-\psi_{2j}^{2})\\
\lambda_j(\psi_{1j}^{2}+\psi_{2j}^{2})-\lambda(\psi_{1j}^{2}-\psi_{2j}^{2})& -2\lambda \psi_{1j}\psi_{2j}\\
 \end{array}\right).$$
{\bf Remark 2.} The system  (\ref{sg10}) is a new integrable
deformation of sine-Gordon equation with the Lax pair given by
(\ref{sg11}). The system  (\ref{sg10}) can be regarded as a new kind
of sine-Gordon equation with self-consistent source which is
different from the sine-Gordon equation with self-consistent source
in \cite{sgws}.

\section{Solutions to  the IDSGE}
\subsection{Multi-soliton solutions }
Introducing the dependent variable transformation
\begin{subequations} \label{transf}
\begin{align} & z=2i\ln\frac{\bar{f}}{f},\label{transfa}\\
& \psi_{1j}=i(\frac{g_j}{f}-\frac{\bar{g_j}}{\bar{f}}),~~~
\psi_{2j}=-(\frac{\bar{g_j}}{\bar{f}}+\frac{g_j}{f}),~~j=1,\cdots,N\label{transfb}
\end{align}
\end{subequations}
the IDSGE (\ref{sg10}) can  be transformed into the bilinear form
\begin{subequations} \label{bilinear}
\begin{align} & D_{y}D_{s}f\cdot f=\frac{1}{2}(f^{2}-\bar{f}^{2})-8i\sum_{j=1}^{N}\lambda_j\bar{g_j}^{2},\label{bilineara}\\
& D_{y}g_j\cdot f=-\lambda_j\bar{g_j}\bar{f},~~j=1,2,\cdots,
N,\label{bilinearb}
\end{align}
\end{subequations}
where $\bar{f}$ and $\bar{g_j}$ are complex conjugates of $f$ and
$g_j$, $D$ is the well-known Hirota bilinear operator defined by
\cite{hirota}
$$D_{y}^{m}D_{s}^{n}f\cdot
g=(\partial_y-\partial_{y'})^{m}(\partial_s-\partial_{s'})^{n}f(y,s)g(y',s')|_{y'=y,s'=s}.$$
The Wronskian determinant is defined as \cite{wr}
\begin{equation}
\label{wronsk}W=|\Psi^{(0)},\Psi^{(1)},\cdots,\Psi^{(N-1)}|=|0,1,\cdots,N-1|=|\widehat{N-1}|,
\end{equation}
where
$\Psi^{(0)}=\Psi=(\Psi_{1}(y,s),\Psi_{2}(y,s),\cdots,\Psi_{N}(y,s))^{T}$
and $\Psi^{(j)}=\frac{\partial^{j} \Psi}{\partial y^{j}}$.

Since the bilinear form (\ref{bilinear}) is same as that
in\cite{sgws} except replacing $g_j$ by $\bar{g}_j$, we may find the
solution of IDSGE (\ref{sg10}) by directly using the formulae and
notation in \cite{sgws}.
We have the following theorem \\
{\bf Theorem 1.} {\sl Let
\begin{equation}
\label{eleme}\Psi_j=ie^{\xi_j}-(-1)^{j}e^{-\xi_j},~~j=1,2,\cdots, N,
\end{equation}
where $\xi_j=-\lambda_j y-\frac{s}{4\lambda_j}+\alpha_j(s)$,
$\lambda_j$ are real number and we set
$\lambda_1<\lambda_2<\cdots<\lambda_N$, then the IDSGE (\ref{sg10})
has the Wronskian determinant solutions given  by  (\ref{transf})
with
\begin{subequations}
\label{wsolu}
\begin{align} & f=|\widehat{N-1}|,\label{wsolua}\\ \nonumber
&
g_h=(-1)^{h+N}\sqrt{\alpha'_h(s)\prod_{l=1}^{h-1}(\lambda_h^{2}-\lambda_l^{2})
\prod_{l=h+1}^{N}(\lambda_l^{2}-\lambda_h^{2})}\overline{|\widehat{N-2},\tau_h|},\\
& \tau_h=(\delta_{h,1},\cdots,\delta_{h,N})^{T},~h=1,\cdots,
N,\label{wsolub}
\end{align}
\end{subequations}
where $\overline{|\widehat{N-2},\tau_h|}$ is  complex conjugates of
$|\widehat{N-2},\tau_h|$. } \\
The Theorem 1 can be proved in the
same way as in \cite{sgws}, we omit it.

When take $N=1$,
 (\ref{transf}) and (\ref{wsolu}) give rise to one soliton solution for the IDSGE(\ref{sg10})
\begin{subequations}
\label{s14}
\begin{align} & z_1=4\arctan e^{2\xi_1}=2i\ln
\frac{1-ie^{2\xi_1}}{1+ie^{2\xi_1}},\label{s14a}\\
& \psi_{11}=\frac{2\sqrt{\alpha'_{1}(s)}e^{3\xi_1}}{1+e^{4\xi_1}},
\psi_{21}=\frac{-2\sqrt{\alpha'_{1}(s)}e^{\xi_1}}{1+e^{4\xi_1}}.\label{s14b}
\end{align}
\end{subequations}
In Fig1, we plot the single-soliton solution of $z_{1}$ and
$\psi_{11}$.
 \vskip 75pt
\begin{center}
\begin{picture}(35,25)
\put(-180,0){\resizebox{!}{3.3cm}{\includegraphics{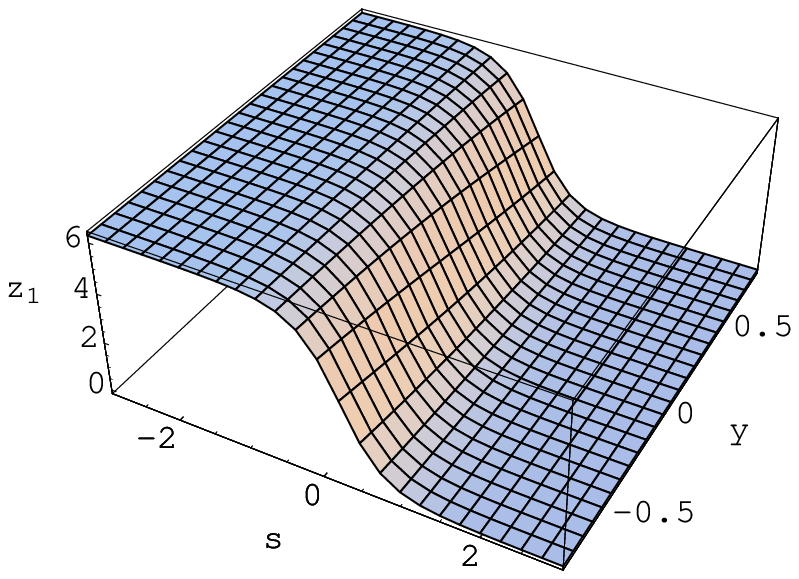}}}
\put(-35,0){\resizebox{!}{3.3cm}{\includegraphics{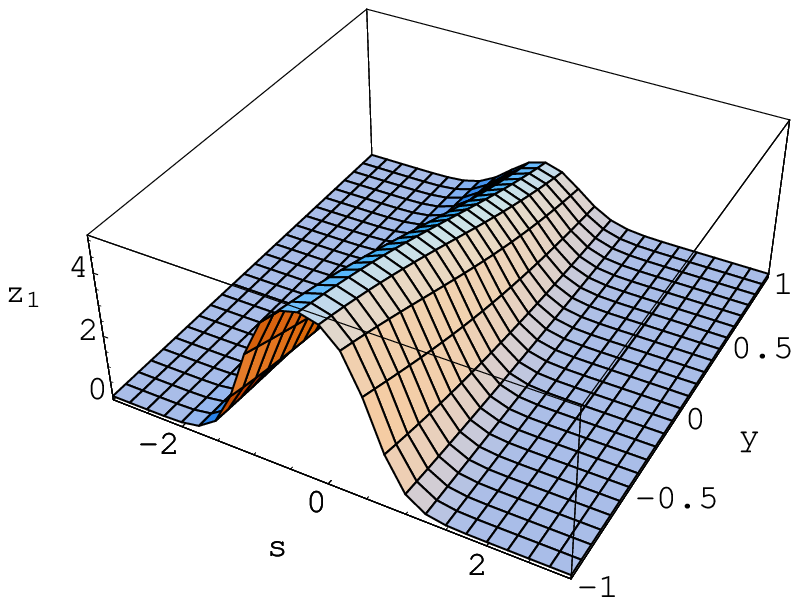}}}
\put(100,0){\resizebox{!}{3.3cm}{\includegraphics{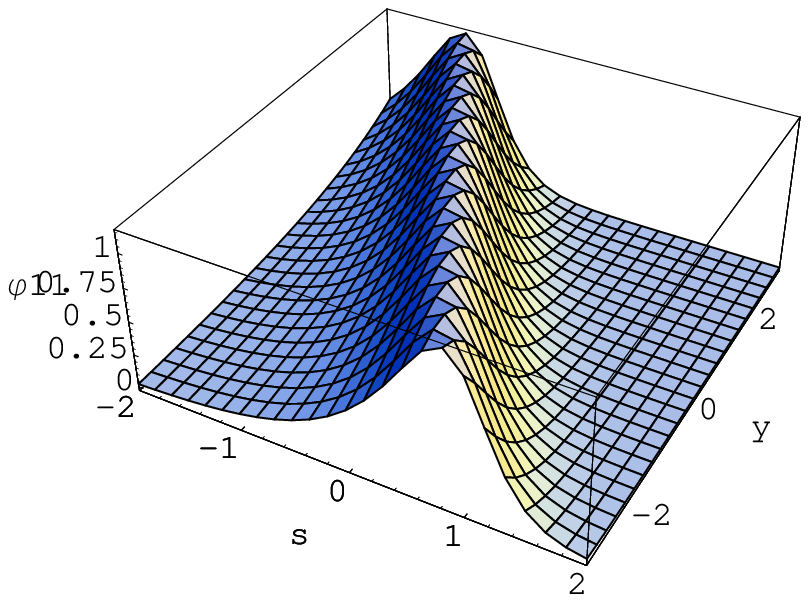}}}
\end{picture}
\end{center}
\begin{center}
\begin{minipage}{ 14cm}{\footnotesize
~~~~~~~~~~~~~~~~(a)~~~~~~~~~~~~~~~~~~~~~~~~~~~~~~~~~~~~~~~~~~~~~~(b)~~~~~~~~~~~~~~~~~~~~~~~~~
~~~~~~~~~~~~~(c)\\
 {\bf Figure 1.} (a),(b) Single soliton solutions $z_{1}$
when $ \lambda_{1}=0.5, ~\alpha_{1}(s)=s,$ and $ \lambda_{1}=0.5,
~\alpha_{1}(s)=s^{2},$ respectively. (c) The eigenfunction
$\psi_{11}$ for $ \lambda_{1}=0.5, ~\alpha_{1}(s)=s.$ }
\end{minipage}
\end{center}
Similarly, when take $N=2$ in (\ref{wsolu}), we have
\begin{subequations}
\label{twosoli}
\begin{align}
&
f=(e^{\xi_{1}+\xi_{2}}+e^{-(\xi_{1}+\xi_{2})})(\lambda_2-\lambda_1)-i(e^{\xi_{1}-\xi_{2}}-e^{\xi_{2}-\xi_{1}})(\lambda_1+\lambda_2),\\
&g_{1}=-\sqrt{\alpha'_1(s)(\lambda_2^{2}-\lambda_1^{2})}(e^{-\xi_{2}}+ie^{\xi_{2}}),~g_{2}=
\sqrt{\alpha'_2(s)(\lambda_2^{2}-\lambda_1^{2})}(e^{-\xi_{1}}-ie^{\xi_{1}}).
\end{align}
\end{subequations}
(\ref{transf}) and (\ref{twosoli}) give to
 the  two
soliton solutions for the IDSGE(\ref{sg10}).
 Fig 2. describes the shapes and
 interactions of two soliton solution for $z_{2}$  and $\psi_{11}$.
Also Fig 2.  shows  the interactions  are elastic collisions and the
influence on two soliton solution for $z_{2}$ and $\psi_{11}$ for
taking different $\alpha_{j}(s)$.

 \vskip 80pt
\begin{center}
\begin{picture}(35,25)
\put(-180,0){\resizebox{!}{3.3cm}{\includegraphics{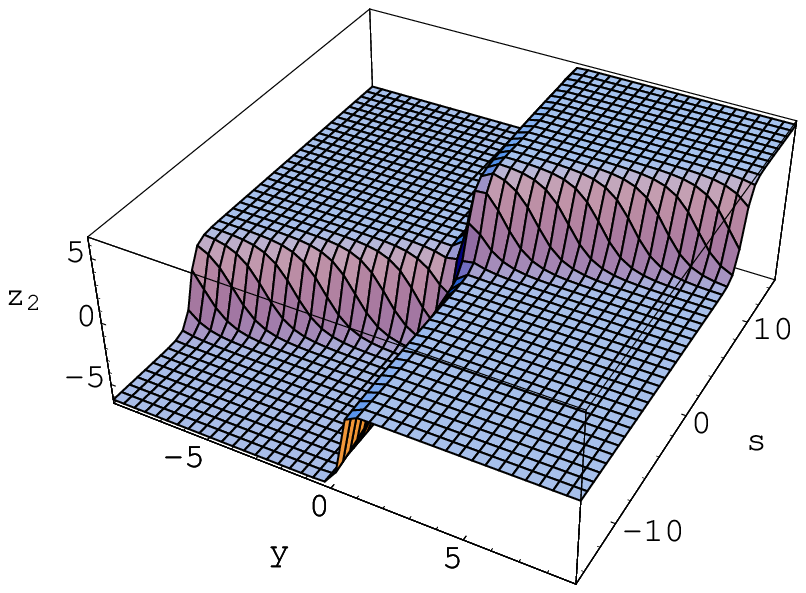}}}
\put(80,0){\resizebox{!}{3.3cm}{\includegraphics{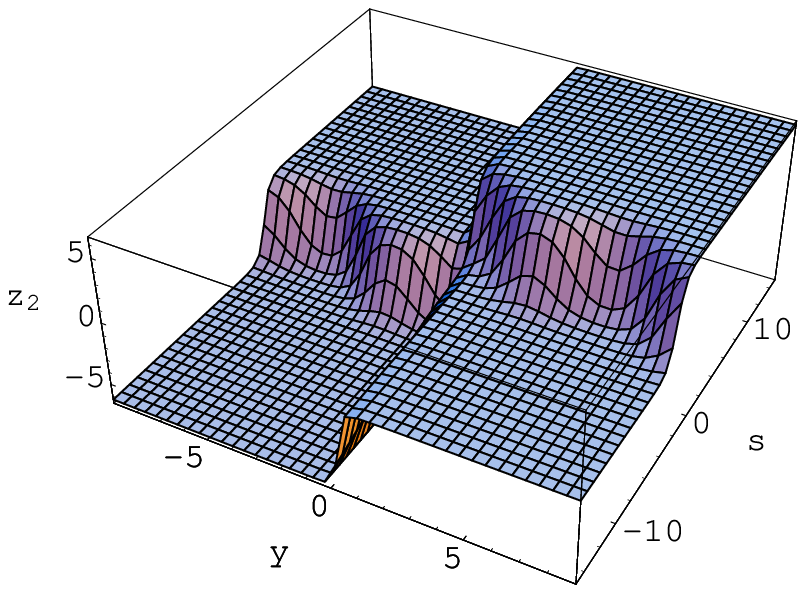}}}
\end{picture}
\end{center}
\begin{center}
\begin{minipage}{ 14cm}{\footnotesize
~~~~~~~~~~~~~~~~~~~~~~(a)~~~~~~~~~~~~~~~~~~~~~~~~~~~~~~~~~~
~~~~~~~~~~~~~~~~~~~~~~~~(b)\\
}
\end{minipage}
\end{center}

\vskip 70pt
\begin{center}
\begin{picture}(35,25)
\put(-150,0){\resizebox{!}{3.3cm}{\includegraphics{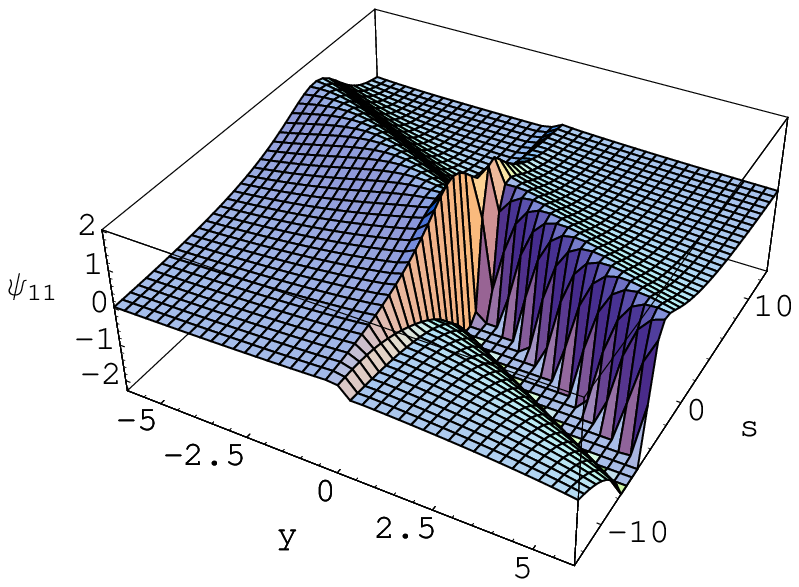}}}
\put(60,0){\resizebox{!}{3.3cm}{\includegraphics{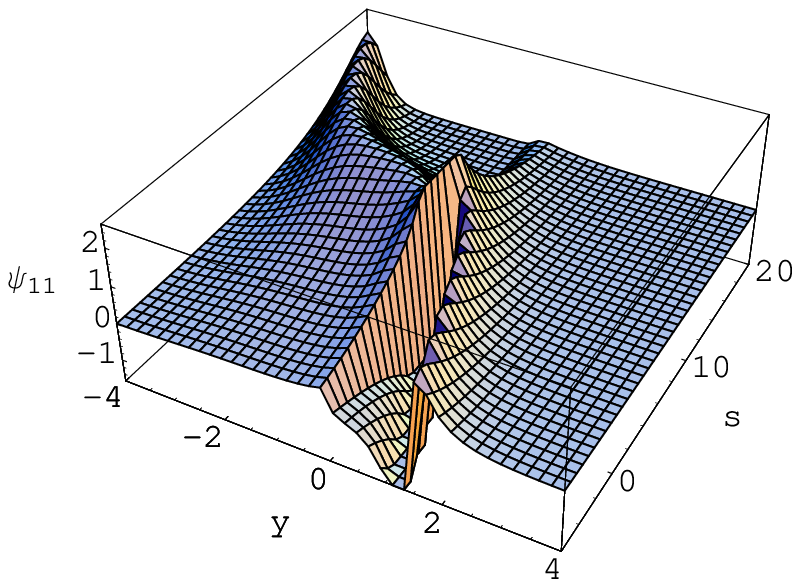}}}
\end{picture}
\end{center}
\begin{center}
\begin{minipage}{ 14cm}{\footnotesize
~~~~~~~~~~~~~~~~~~~~~~~~(c)~~~~~~~~~~~~~~~~~~~~~~~~~~~~~~~~~~~~~~~~~~~~~~
~~~~~~~~~~~~~(d)\\
{\bf Figure 2.}(a),(b) The shapes and interactions  for the two
soliton solutions $z_{2}$ when $\lambda_{1}=-0.1,~\lambda_{2}=1,~
\alpha_{1}(s)=s,~\alpha_{2}(s)=s$ and
$\lambda_{1}=-0.1,~\lambda_{2}=1,~
\alpha_{1}(s)=\sin2s,~\alpha_{2}(s)=\cos s,$ respectively. (c),(d)
The shapes and interactions for $\psi_{11}$  when
$\lambda_{1}=-0.1,~\lambda_{2}=0.2,~
\alpha_{1}(s)=2s,~\alpha_{2}(s)=s,$ and
$\lambda_{1}=-0.1,~\lambda_{2}=0.2,~
\alpha_{1}(s)=4s,~\alpha_{2}(s)=\cos s,$ respectively.}
\end{minipage}
\end{center}
 Notice that solutions (\ref{wsolu}) contains
arbitrary $s$ functions $\alpha_{j}(s)$. This implies that the
insertion of non-homogeneous terms into the soliton equation may
cause the variation of the speed and shape of soliton. So the
dynamics of solutions of IDSGE(\ref{sg10}) turns out to be much
richer than that of solutions of sine-Gordon equation.

\subsection{Negaton solutions and positon solutions }
For $N=2$, taking
$\alpha_1(s)=c_1,~\alpha_2(s)=(\lambda_2-\lambda_1)e(s)+c_1-\frac{1}{2}i\pi$,
then (\ref{eleme}) leads to
\begin{equation}
\label{n1}
\Psi_1=e^{-\xi_{1}}+ie^{\xi_{1}},~\Psi_2=-i(e^{-\xi_{2}}+ie^{\xi_{2}}),
\end{equation}
where
$\xi_{1}=-\lambda_1y-\frac{s}{4\lambda_1}+c_1,~\xi_{2}=-\lambda_2y-\frac{s}{4\lambda_2}+(\lambda_2-\lambda_1)e(s)+c_1-\frac{1}{2}i\pi,$
 $c_1$ is a constant and $e(s)$ is a function for $s$. We have
\begin{subequations}
\label{n2}
\begin{align}
&f=\left|\begin{array}{cc}
\Psi_1 &\Psi_{1y}\\
 \Psi_{2}&\Psi_{2y}\\
 \end{array}\right|=\left|\begin{array}{cc}
\Psi_1 &\Psi_{1y}\\
\frac{\partial \Psi_2}{\partial \lambda_2}|_{\lambda_2=\lambda_1} &\frac{\partial^{2} \Psi_2}{\partial \lambda_2\partial y}|_{\lambda_2=\lambda_1}\\
 \end{array}\right|(\lambda_2-\lambda_1)+o(\lambda_2-\lambda_1)\nonumber\\
 &~~~~~=-(e^{2\xi_{1}}+e^{-2\xi_{1}}+4i\lambda_1\gamma)(\lambda_2-\lambda_1)+o(\lambda_2-\lambda_1),\\
& g_1=0,\\
&g_2=\sqrt{(\lambda_2-\lambda_1)e'(s)(\lambda_2^{2}-\lambda_1^{2})}\overline{\left|\begin{array}{cc}
\Psi_1 &0\\
\Psi_2 &1\\
 \end{array}\right|}=(\lambda_2-\lambda_1)\sqrt{e'(s)(\lambda_2+\lambda_1)}(e^{-\xi_{1}}-ie^{\xi_{1}}),
\end{align}
\end{subequations}
where $\gamma=y-\frac{s}{4\lambda_1^{2}}-e(s)$. Then we obtain the
one negaton solution from (\ref{transf}) by taking
$\lambda_2\rightarrow \lambda_1$
\begin{subequations}
\label{n3}
\begin{align}
&z=2i\ln \frac{ch2\xi_{1}-2i\lambda_1\gamma}{ch2\xi_{1}+2i\lambda_1\gamma},\\
&\psi_{12}=\frac{2\sqrt{2\lambda_1e'(s)}(-4\lambda_1\gamma
e^{-\xi_{1}}+e^{-\xi_{1}}+e^{3\xi_{1}})}{(e^{2\xi_{1}}+e^{-2\xi_{1}})^{2}+16\lambda_1^{2}\gamma^{2}},\\
&\psi_{22}=\frac{2\sqrt{2\lambda_1e'(s)}(4\lambda_1\gamma
e^{\xi_{1}}+e^{\xi_{1}}+e^{-3\xi_{1}})}{(e^{2\xi_{1}}+e^{-2\xi_{1}})^{2}+16\lambda_1^{2}\gamma^{2}}.
\end{align}
\end{subequations}

 \vskip 80pt
\begin{center}
\begin{picture}(35,25)
\put(-150,0){\resizebox{!}{3.3cm}{\includegraphics{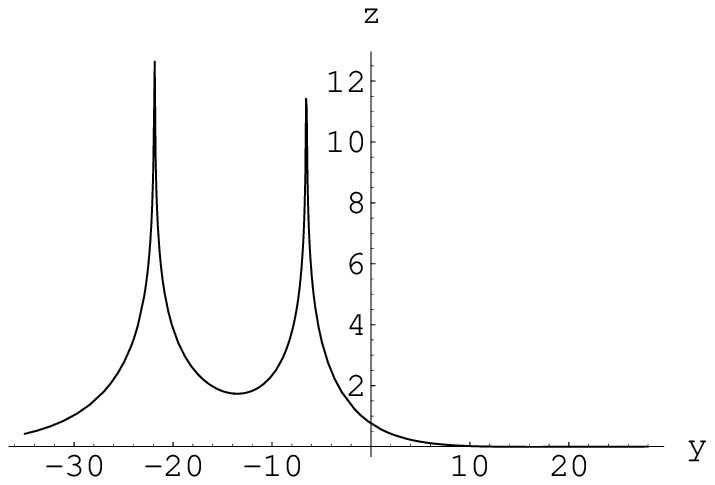}}}
\put(40,0){\resizebox{!}{3.3cm}{\includegraphics{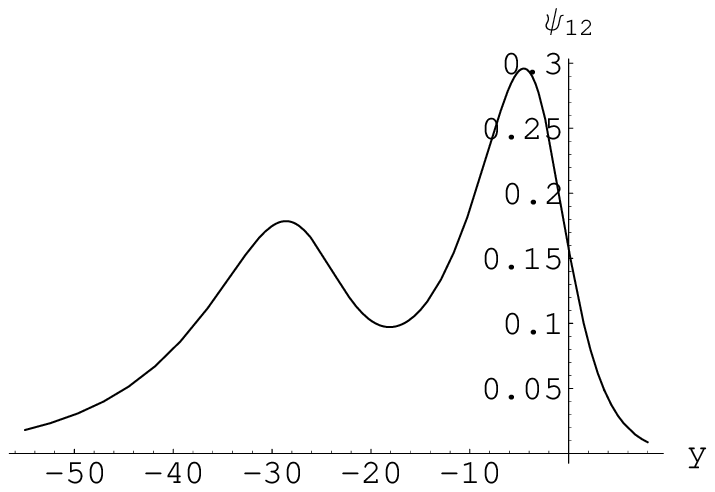}}}
\end{picture}
\end{center}
\begin{center}
\begin{minipage}{ 16cm}{\footnotesize
~~~~~~~~~~~~~~~~~~~~~~~~~~~~(a)~~~~~~~~~~~~~~~~~~~~~~~~~~~~~~~~~~~~~~~~~~~~~~
~~~~~~~~~~~~~(b)\\
{\bf Figure 3.} The shapes for one negaton solution $z$ and
$\psi_{12}$ when $\lambda_{1}=0.1,~ e(s)=2s,~s=0.5$. }
\end{minipage}
\end{center}
The shapes are given in Fig.3. In general, as proposed in
\cite{mat}, the $N$ negaton solution can be obtained from
(\ref{transf}), (\ref{wronsk}) and  (\ref{wsolu}) by replacing $N$
by $2N$, taking
\begin{subequations}
\label{np}
\begin{align}
&\Psi^{(0)}=(\Psi_{1},\frac{\partial \Psi_{2}}{\partial \lambda_2}|_{\lambda_2=\lambda_1},\Psi_{3},
\frac{\partial \Psi_{4}}{\partial \lambda_4}|_{\lambda_4=\lambda_3}\cdots,\Psi_{2N-1},
\frac{\partial \Psi_{2N}}{\partial \lambda_{2N}}|_{\lambda_{2N}=\lambda_{2N-1}})^{T},\\
& \xi_{2k-1}=-\lambda_{2k-1}y-\frac{s}{4\lambda_{2k-1}}+c_{2k-1},\\
&\xi_{2k}=-\lambda_{2k}y-\frac{s}{4\lambda_{2k}}+(\lambda_{2k}-\lambda_{2k-1})e_{2k}(s)+c_{2k-1}-\frac{1}{2}i\pi,
\end{align}
\end{subequations}
and taking $\lambda_{2k}\rightarrow\lambda_{2k-1}$.

In order to derive the positon solution as pointed out in
\cite{beu}, we have to take
\begin{equation}
\label{p1} \lambda_1=i\mu_1,~\lambda_2=i\mu_2,~c_1=-i\bar{c}_1.
\end{equation}
By similarly calculation, we obtain the following one positon
solution
\begin{subequations}
\label{n4}
\begin{align}
&z=2i\ln \frac{\cos2\eta_{1}+2\mu_1\bar{\gamma}}{\cos2\eta_{1}-2\mu_1\bar{\gamma}},\\
&\psi_{12}=\frac{\sqrt{2i\mu_1e'(s)}(-4i\mu_1\bar{\gamma}
e^{i\eta_{1}}+e^{i\eta_{1}}+e^{-3i\eta_{1}})}{2(\cos^{2}2\eta_{1}-4\mu_1^{2}\bar{\gamma}^{2})},\\
&\psi_{22}=\frac{\sqrt{2i\mu_1e'(s)}(4i\mu_1\bar{\gamma}
e^{-i\eta_{1}}+e^{-i\eta_{1}}+e^{3i\eta_{1}})}{2(\cos^{2}2\eta_{1}-4\mu_1^{2}\bar{\gamma}^{2})}.
\end{align}
\end{subequations}
where
$\bar{\gamma}=y+\frac{s}{4\mu_1^{2}}-e(s),~\eta_1=\mu_1y-\frac{s}{4\mu_1}+\bar{c}_1$.
Just as the positon solutions of the sine-Gordon equation are
complex\cite{beu}, one positon solution (\ref{n4}) is also complex
and reduces to the positon solution of sine-Gordon equation by
taking $e(s)$ to be a constant.

 \vskip 80pt
\begin{center}
\begin{picture}(35,25)
\put(-210,0){\resizebox{!}{3.3cm}{\includegraphics{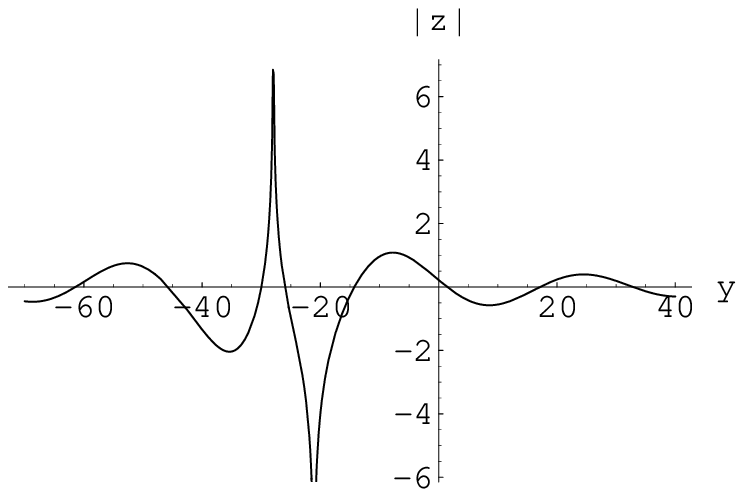}}}
\put(-60,0){\resizebox{!}{3.3cm}{\includegraphics{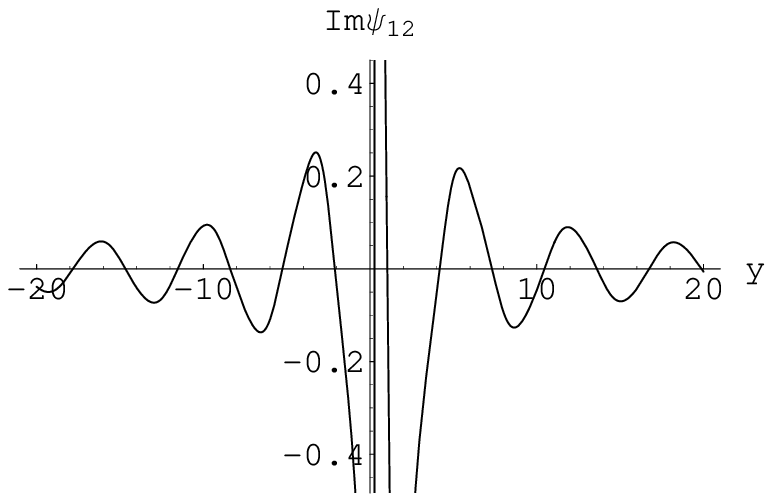}}}
\put(100,0){\resizebox{!}{3.3cm}{\includegraphics{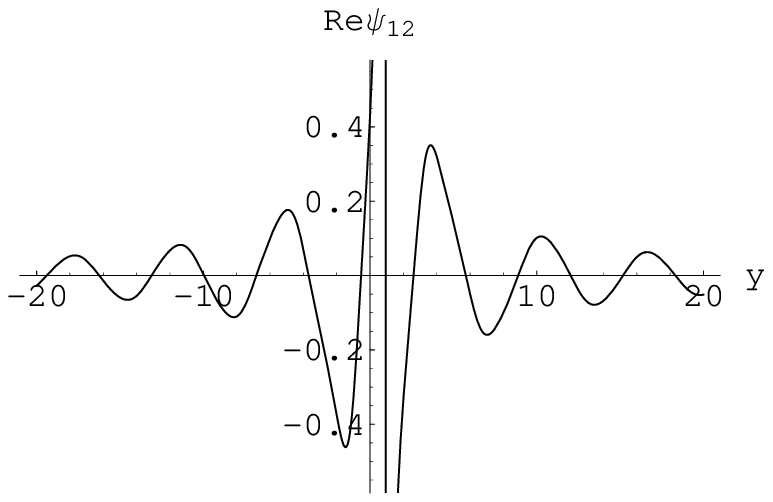}}}
\end{picture}
\end{center}
\begin{center}
\begin{minipage}{ 14cm}{\footnotesize
~~~~~~(a)~~~~~~~~~~~~~~~~~~~~~~~~~~~~~~~~~~~~~~~~~~~~~(b)~~~~
~~~~~~~~~~~~~~~~~~~~~~~~~~~~~~~~~~~~~(c)\\
{\bf Figure 4.} The shapes for the modulus of one positon solution
$z$, the real part and imaginary part of $\psi_{12}$ when
$\lambda_{1}=0.1,~ e(s)=s^{2},~s=1$, respectively. }
\end{minipage}
\end{center}
Fig.4 describes the shapes for the modulus of one positon solution.
Similarly, we can find $N$ positon solutions by using (\ref{np}) and
$\lambda_j=i\mu_j,~c_k=-i\bar{c}_k$ and $\mu_{2k}\rightarrow
\mu_{2k-1}$.

\section{Solutions for the IDSPE}
\subsection{N-loop soliton solutions}

{\bf Proposition 1.} {\sl Suppose that $z$ and
$\psi_{ij},~(i=1,2,~j=1,2,\cdots, N)$ are solutions of the IDSGE
(\ref{sg10}), then the solutions of the IDSPE (\ref{sp14}) with a
parametric representation of (y,s) are given by
\begin{subequations}
\label{s22}
\begin{align}
&u=z_{s}-2\sum_{j=1}^{N}(\psi_{1j}^{2}+\psi_{2j}^{2}),\label{s22a}\\
&\varphi_{1j}(x,t)=\psi_{1j}(y,s),~\varphi_{2j}(x,t)=\psi_{2j}(y,s),~j=1,2,\cdots,N,\label{s22b}\\
&x(y,s)=\int \cos zdy=y-2(\ln
f\bar{f}|_{\alpha_j(s)=\alpha_j})_{s}|_{\alpha_j=\alpha_j(s)},\label{s22c}
\end{align}
\end{subequations}
where $~\alpha_j$ are arbitrary
constants.} \\
{\bf Proof.} It is obviously that (\ref{s22a}) is given by
(\ref{sgad1}). In the following, we prove (\ref{s22c}).
 From the
reciprocal transformation, we have the following linear PDEs for $x$
\begin{equation}
\label{s20} \frac{\partial x}{\partial y}=\frac{1}{r},~
\frac{\partial x}{\partial
s}=-\frac{1}{2}u^{2}-2\sum_{j=1}^{N}\lambda_j^{-1}\psi_{1j}\psi_{2j}.
\end{equation}
By making use of the compatibility of the above two equations, we
have
\begin{equation}
\label{s21} x(y,s)=\int\frac{1}{r}dy=\int cos zdy.
\end{equation}
From (\ref{transf}) and (\ref{bilinear}), a direct calculation gives
\begin{equation}
\label{rela1} cos z=1-2(\ln
f\bar{f})_{ys}+8i\sum_{j=1}^{N}\lambda_j(\frac{g_j^{2}}{\bar{f}^{2}}-\frac{\bar{g}_j^{2}}{f^{2}}).
\end{equation}
When $f$ and $g_j$ are given by (\ref{eleme}) and  (\ref{wsolu}),
the terms with $\alpha'(s)$ in $2(\ln f\bar{f})_{ys}$  and the terms
$8i\sum_{j=1}^{N}\lambda_j(\frac{g_j^{2}}{\bar{f}^{2}}-\frac{\bar{g}_j^{2}}{f^{2}})$
are canceled. So the  above equation becomes
\begin{equation}
\label{rela2} cos z=1-2(\ln
f\bar{f}|_{\alpha_j(s)=\alpha_j})_{ys}|_{\alpha_j=\alpha_j(s)},
\end{equation}
namely, for calculating  the derivatives with respect to $s$, we
regard $\alpha_j(s)$ as independent of $s$. Substituting this
equation in (\ref{s21}) leads to (\ref{s22c}).

For example, when take $N=1$, by making use of  (\ref{s14}) and
(\ref{s22}), we obtain the one loop soliton solution for the IDSPE
(\ref{sp14})
\begin{subequations}
\label{s23}
\begin{align} & u_1=\frac{2e^{2\xi_1}}{\lambda_1(1+e^{4\xi_1})},\label{s23a}\\
&
\varphi_{11}=\frac{2\sqrt{\alpha'_{1}(s)}e^{3\xi_1}}{(1+e^{4\xi_1})},~~
\varphi_{21}=\frac{-2\sqrt{\alpha'_{1}(s)}e^{\xi_1}}{(1+e^{4\xi_1})}.\label{s23b}\\
&x(y,s)=y+\frac{2}{\lambda_1(1+e^{4\xi_1})}.
\end{align}
\end{subequations}

\vskip 80pt
\begin{center}
\begin{picture}(35,25)
\put(-190,0){\resizebox{!}{3.3cm}{\includegraphics{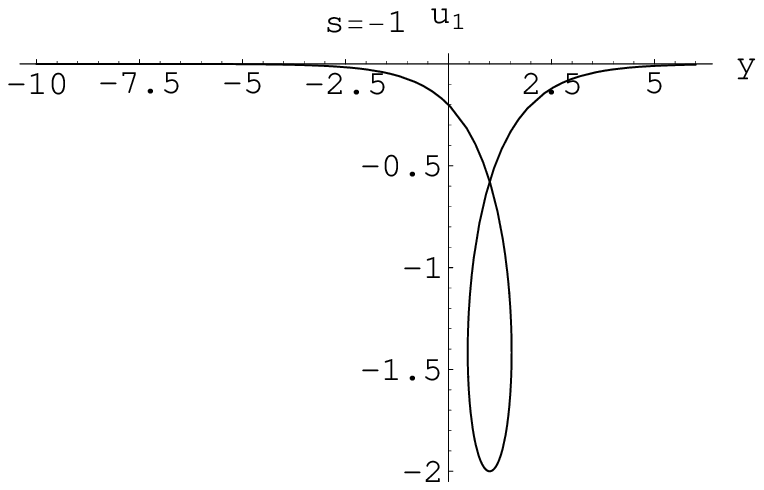}}}
\put(-40,0){\resizebox{!}{3.3cm}{\includegraphics{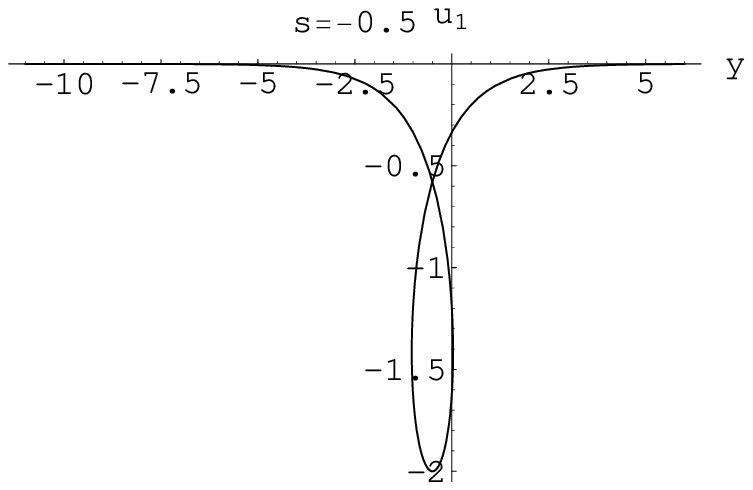}}}
\put(120,0){\resizebox{!}{3.3cm}{\includegraphics{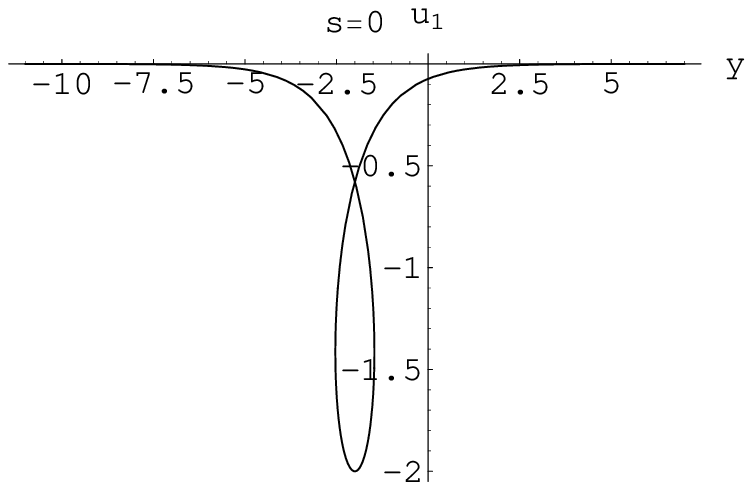}}}
\end{picture}
\end{center}
\begin{center}
\begin{minipage}{ 14cm}{\footnotesize
~~~~~~~~~~~~~~~~~~~~~~(a)~~~~~~~~~~~~~~~~~~~~~~~~~~~~~~~~~~~~~(b)~~~~
~~~~~~~~~~~~~~~~~~~~~~~~~~~~~~~~~~~~~~~~~(c)\\
}
\end{minipage}
\end{center}

\vskip 80pt
\begin{center}
\begin{picture}(35,25)
\put(-190,0){\resizebox{!}{3.3cm}{\includegraphics{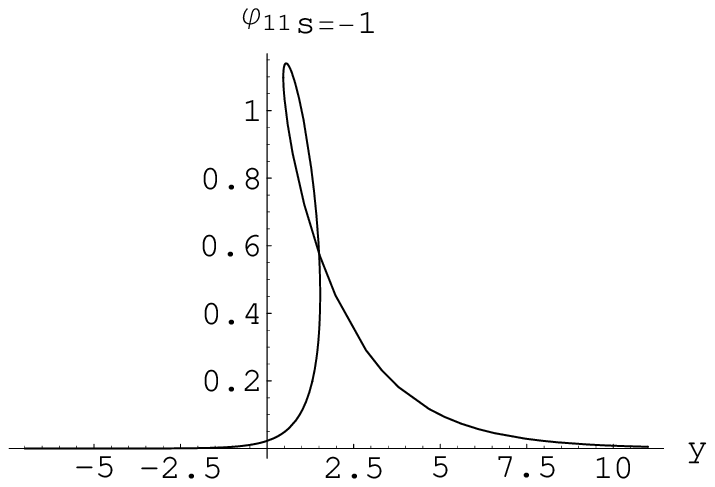}}}
\put(-40,0){\resizebox{!}{3.3cm}{\includegraphics{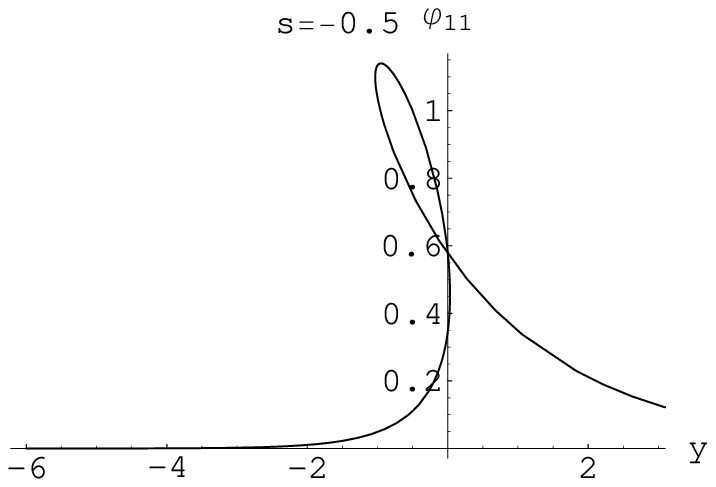}}}
\put(120,0){\resizebox{!}{3.3cm}{\includegraphics{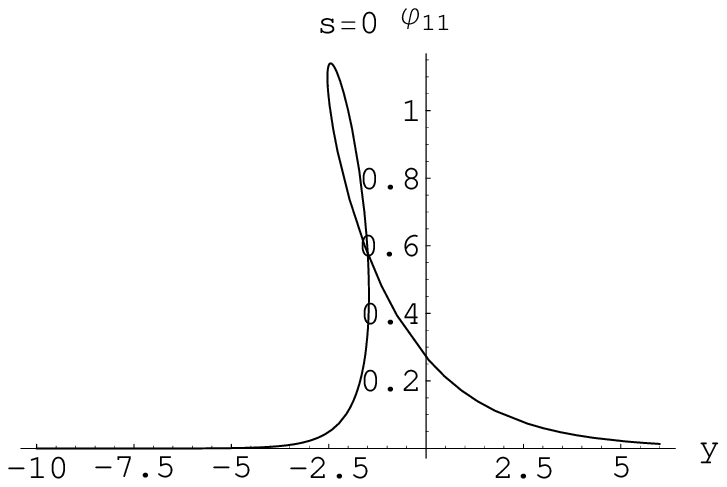}}}
\end{picture}
\end{center}
\begin{center}
\begin{minipage}{ 14cm}{\footnotesize
~~~~~~~~~~~~~~~~~~~~~~(d)~~~~~~~~~~~~~~~~~~~~~~~~~~~~~~~~~~(e)~~~~~~~
~~~~~~~~~~~~~~~~~~~~~~~~~~~~~~~~~~~~~~~~~(f)\\
{\bf Figure 5.}  The shapes and motions for the one loop soliton
solutions $u_{1}$ and $\varphi_{11}$ when
$\lambda_{1}=-0.5,~\alpha_{1}(s)=s$.}
\end{minipage}
\end{center}
Fig.5. shows that the one loop soliton solutions move to the left
and keep the shapes. When take $N=2$ in (\ref{wsolu}),
(\ref{transf}) and (\ref{s22}) give to the two loop solution of the
IDSPE (\ref{sp14})
\begin{subequations}
\label{twoloop}
\begin{align}
&u_2=\frac{2(\lambda_{1}^{2}-\lambda_{2}^{2})[\lambda_{1}(1+e^{4\xi_1})e^{2\xi_2}(1-12\lambda_{2}\alpha'_2(s))
-\lambda_{2}(1+e^{4\xi_2})e^{2\xi_1}(1-12\lambda_{1}\alpha'_1(s))]}{\lambda_{1}\lambda_{2}[(1+e^{4\xi_1})(1+e^{4\xi_2})
(\lambda_{1}^{2}+\lambda_{2}^{2})-2\lambda_{1}\lambda_{2}(1-e^{4\xi_1}-e^{4\xi_2}+4e^{2(\xi_1+\xi_2)}+e^{4(\xi_1+\xi_2)})]},\label{twoloopa}\\
&\varphi_{11}=\frac{4\sqrt{(\lambda_{2}^{2}-\lambda_{1}^{2})\alpha'_1(s)}[(1-e^{4\xi_2}+2e^{2(\xi_1+\xi_2})\lambda_{1}-
(1+e^{4\xi_2})\lambda_{1}]e^{\xi_1}}{(1+e^{4\xi_1})(1+e^{4\xi_2})
(\lambda_{1}^{2}+\lambda_{2}^{2})-2\lambda_{1}\lambda_{2}(1-e^{4\xi_1}-e^{4\xi_2}+4e^{2(\xi_1+\xi_2)}+e^{4(\xi_1+\xi_2)})},\label{twoloopb}\\
&\varphi_{21}=\frac{4\sqrt{(\lambda_{2}^{2}-\lambda_{1}^{2})\alpha'_1(s)}[(e^{2\xi_2}-e^{2\xi_1}+e^{2\xi_1+4\xi_2})\lambda_{1}-
e^{2\xi_1}(1+e^{4\xi_2})\lambda_{2}]e^{\xi_1}}{(1+e^{4\xi_1})(1+e^{4\xi_2})
(\lambda_{1}^{2}+\lambda_{2}^{2})-2\lambda_{1}\lambda_{2}(1-e^{4\xi_1}-e^{4\xi_2}+4e^{2(\xi_1+\xi_2)}+e^{4(\xi_1+\xi_2)})},\label{twoloopc}\\
&\varphi_{12}=\frac{4\sqrt{(\lambda_{2}^{2}-\lambda_{1}^{2})\alpha'_2(s)}[(2e^{2\xi_1}-e^{2\xi_2}+e^{4\xi_1+2\xi_2})\lambda_{2}-
e^{2\xi_2}(1+e^{4\xi_1})\lambda_{1}]e^{\xi_2}}{(1+e^{4\xi_1})(1+e^{4\xi_2})
(\lambda_{1}^{2}+\lambda_{2}^{2})-2\lambda_{1}\lambda_{2}(1-e^{4\xi_1}-e^{4\xi_2}+4e^{2(\xi_1+\xi_2)}+e^{4(\xi_1+\xi_2)})},\\
&\varphi_{22}=\frac{4\sqrt{(\lambda_{2}^{2}-\lambda_{1}^{2})\alpha'_2(s)}[(1-e^{4\xi_1}+2e^{2(\xi_1+\xi_2)})\lambda_{2}-
(1+e^{4\xi_1})\lambda_{1}]e^{\xi_2}}{(1+e^{4\xi_1})(1+e^{4\xi_2})
(\lambda_{1}^{2}+\lambda_{2}^{2})-2\lambda_{1}\lambda_{2}(1-e^{4\xi_1}-e^{4\xi_2}+4e^{2(\xi_1+\xi_2)}+e^{4(\xi_1+\xi_2)})}.
\end{align}
\end{subequations}

Similarly, by making use of (\ref{eleme}), (\ref{wsolu}) and
(\ref{s22}), we can obtain the N-loop soliton solution for the IDSPE
(\ref{sp14}). Fig. 6. describes the shapes and interactions of the
two loop soliton solutions and  the interactions of two loop soliton
solution for $u_{2}$  and $\varphi_{11}$ are shown to be elastic
collisions. \vskip 80pt
\begin{center}
\begin{picture}(35,25)
\put(-190,0){\resizebox{!}{3.3cm}{\includegraphics{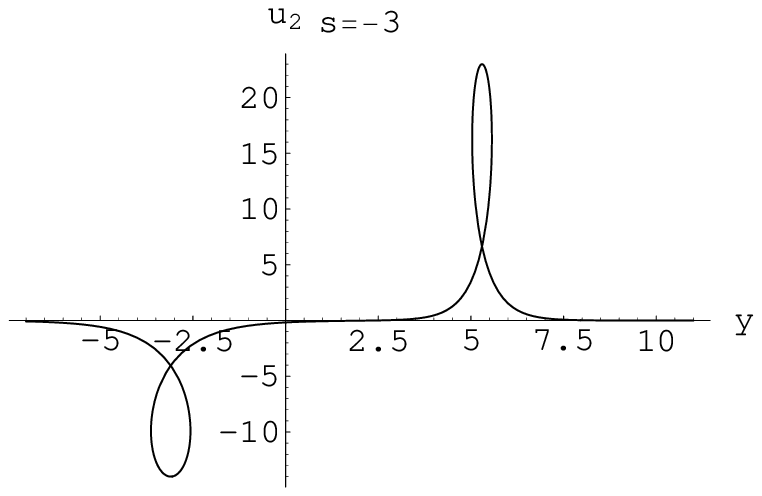}}}
\put(-40,0){\resizebox{!}{3.3cm}{\includegraphics{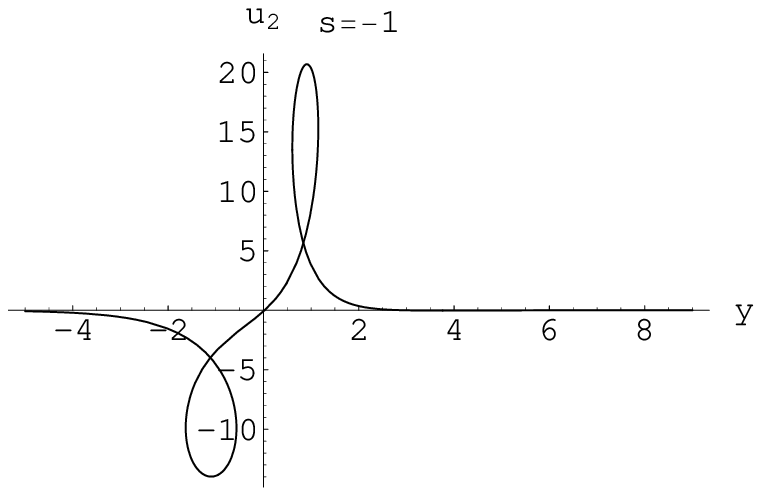}}}
\put(120,0){\resizebox{!}{3.3cm}{\includegraphics{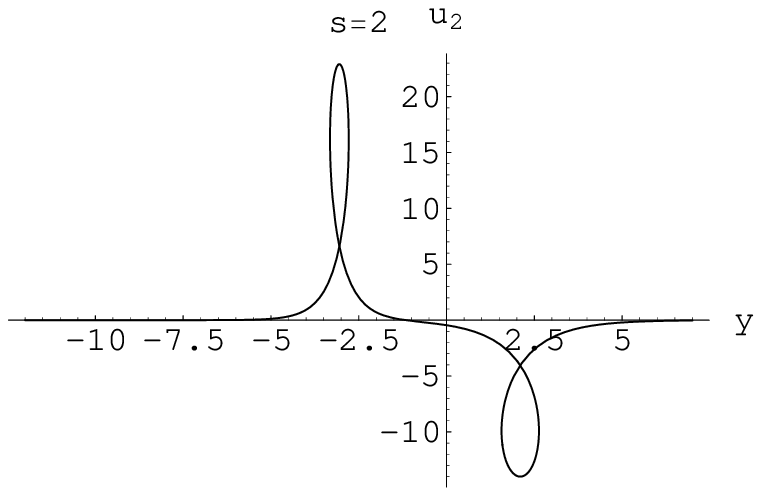}}}
\end{picture}
\end{center}
\begin{center}
\begin{minipage}{ 14cm}{\footnotesize
~~~~~~~~~~~~~~~~~~~~~~(a)~~~~~~~~~~~~~~~~~~~~~~~~~~~~~~~~~~~~~(b)~~~~
~~~~~~~~~~~~~~~~~~~~~~~~~~~~~~~~~~~~~~~~~(c)\\
}
\end{minipage}
\end{center}

\vskip 80pt
\begin{center}
\begin{picture}(35,25)
\put(-190,0){\resizebox{!}{3.3cm}{\includegraphics{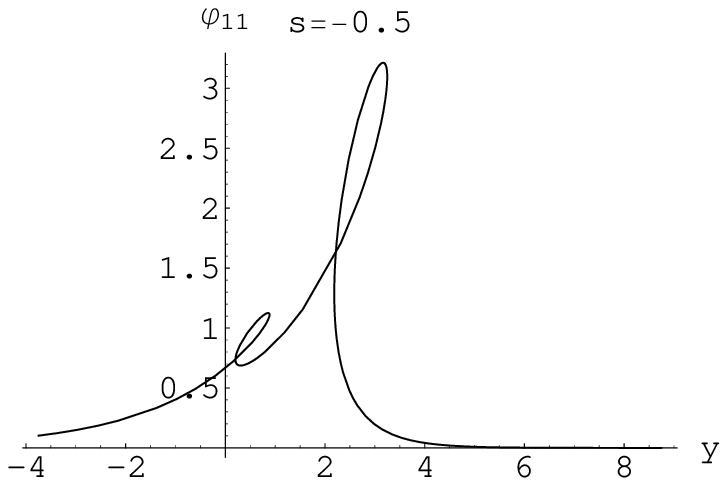}}}
\put(-40,0){\resizebox{!}{3.3cm}{\includegraphics{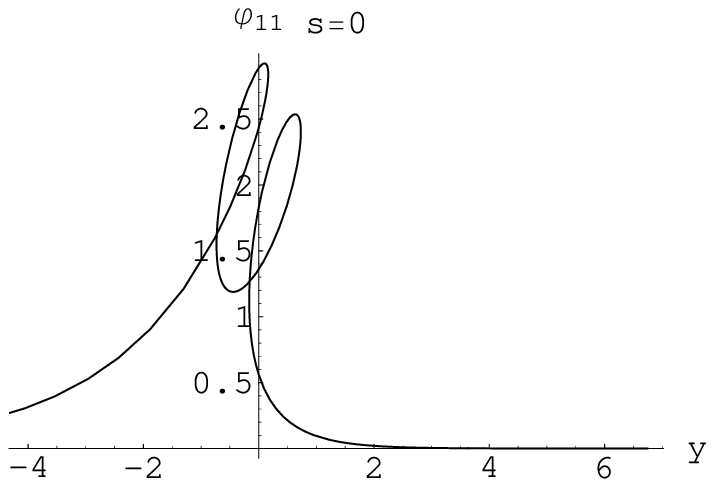}}}
\put(120,0){\resizebox{!}{3.3cm}{\includegraphics{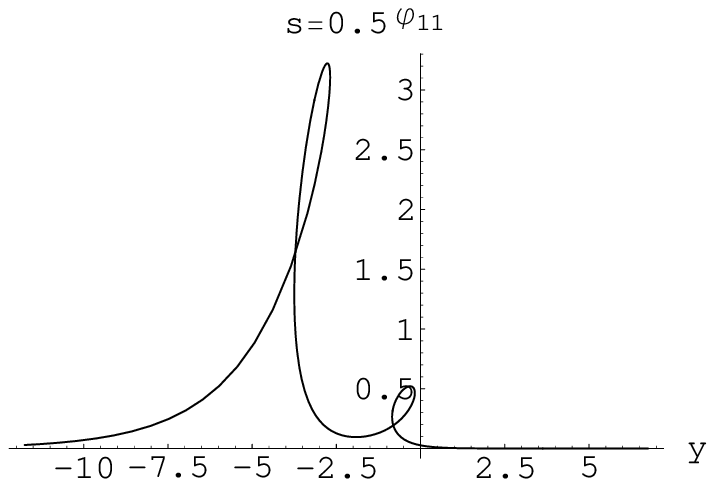}}}
\end{picture}
\end{center}
\begin{center}
\begin{minipage}{ 14cm}{\footnotesize
~~~~~~~~~~~~~~~~~~~~~~(d)~~~~~~~~~~~~~~~~~~~~~~~~~~~~~~~~~~(e)~~~~~~~
~~~~~~~~~~~~~~~~~~~~~~~~~~~~~~~~~~~~~~~~~(f)\\
{\bf Figure 6.}  The shapes and interactions for the two loop
soliton solutions $u_{2}$ and $\varphi_{11}$ when
$\lambda_{1}=-1,~\lambda_{2}=0.5,~\alpha_{1}(s)=2s,~\alpha_{2}(s)=s$.}
\end{minipage}
\end{center}

\subsection{Negaton solutions and positon solutions}
By making use of  (\ref{transf}),~(\ref{n3}) and (\ref{s22}), the
one negaton solution for the IDSPE (\ref{sp14}) is given
\begin{subequations}
\label{nega}
\begin{align} & u=\frac{-2e^{\xi_1}[(s-4\lambda_1^{2}(y-e(s)))(e^{4\xi_1}-1)+2\lambda_1(1+8\lambda_1^{2}e'(s))(e^{4\xi_1}+1)]}
{[s^{2}+16\lambda_1^{4}(y-e(s))^{2}+2\lambda_1^{2}(1-4ys)-8\lambda_1^{2}se(s)]e^{4\xi_1}+\lambda_1^{2}(e^{8\xi_1}+1)},\label{s23a}\\
&
\varphi_{12}=\sqrt{2\lambda_1e'(s)}\frac{e^{-\xi_1}+e^{3\xi_1}-4\lambda_1\gamma
e^{-\xi_1}}{2(ch^{2}2\xi_1+4\lambda_1^{2}\gamma^{2})},~
\varphi_{22}=\sqrt{2\lambda_1e'(s)}\frac{e^{\xi_1}+e^{-3\xi_1}+4\lambda_1\gamma
e^{\xi_1}}{2(ch^{2}2\xi_1+4\lambda_1^{2}\gamma^{2})},\label{s23b}\\
&x(y,s)=y-\frac{2\{[2s+32\lambda_1^{4}(y+e(s))e'(s)-8\lambda_1^{2}(y+e(s)+se'(s))]e^{-4\xi_1}+\lambda_1(e^{-8\xi_1}-1)\}}
{[s^{2}+16\lambda_1^{4}(y-e(s))^{2}+2\lambda_1^{2}(1-4ys)+8se(s)]e^{-4\xi_1}+\lambda_1^{2}(e^{-8\xi_1}+1)}.
\end{align}
\end{subequations}

\vskip 80pt
\begin{center}
\begin{picture}(35,25)
\put(-150,0){\resizebox{!}{3.3cm}{\includegraphics{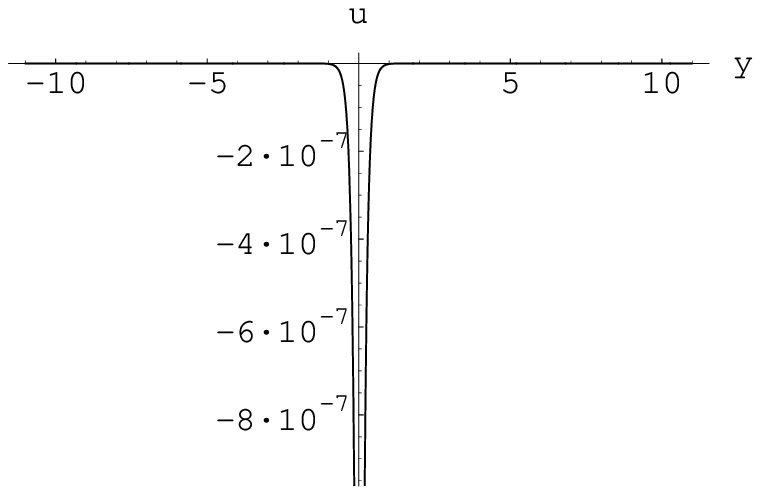}}}
\put(80,0){\resizebox{!}{3.3cm}{\includegraphics{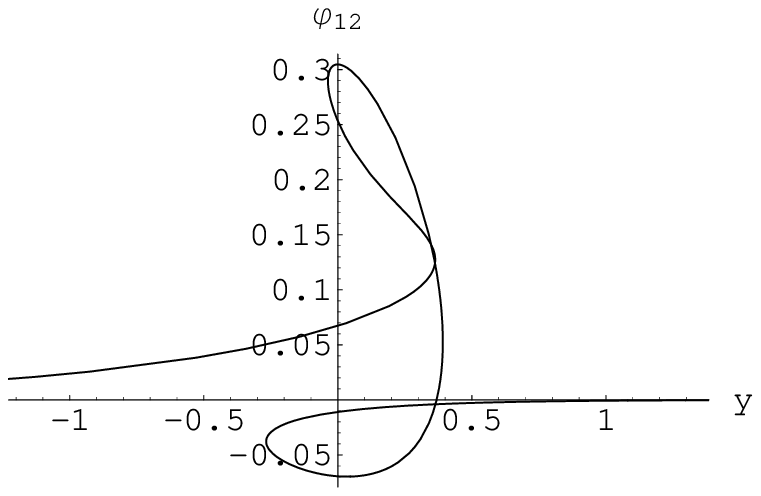}}}
\end{picture}
\end{center}
\begin{center}
\begin{minipage}{ 14cm}{\footnotesize
~~~~~~~~~~~~~~~~~~~~~~(a)~~~~~~~~~~~~~~~~~~~~~~~~~~~~~~~~~~~~~~~~
~~~~~~~~~~~~~~~~~~~~~~~~~~~~~~~~~~~(b)\\
{\bf Figure 7.}  The shapes for the one negaton   solutions  when
$\lambda_{1}=1,~e(s)=s^{3},~s=-0.2 $.}
\end{minipage}
\end{center}
The shapes of the one negaton   solutions are given in Fig.7. In the
same way, by making use of (\ref{transf}),~(\ref{n4}) and
(\ref{s22}), the one positon solution for the IDSPE (\ref{sp14}) is
given

\begin{subequations}
\label{posi}
\begin{align} & u=\frac{4i[(s+4\mu_1^{2}(y-e(s)))\sin2\eta_1-2\mu_1(1-8\mu_1^{2}e'(s))\cos2\eta_1]}{s^{2}+16\mu_1^{4}
(y+e(s))^{2}-2\mu_1^{2}(1-4sy+\cos4\eta_1-4se(s))},\label{posia}\\
& \varphi_{12}=\frac{\sqrt{2i\mu_1e'(s)}(-4i\mu_1\bar{\gamma}
e^{i\eta_{1}}+e^{i\eta_{1}}+e^{-3i\eta_{1}})}{2(\cos^{2}2\eta_{1}-4\mu_1^{2}\bar{\gamma}^{2})},\\
&\varphi_{22}=\frac{\sqrt{2i\mu_1e'(s)}(4i\mu_1\bar{\gamma}
e^{-i\eta_{1}}+e^{-i\eta_{1}}+e^{3i\eta_{1}})}{2(\cos^{2}2\eta_{1}-4\mu_1^{2}\bar{\gamma}^{2})}.\label{posib}\\
&x(y,s)=y-\frac{4[s+4\mu_1^{2}(y-e(s))-\sin4\eta_1]}{s^{2}+16\mu_1^{4}(y-e(s))^{2}-2\mu_1^{2}(1-4s(y-e(s))+\cos4\eta_1)}.
\end{align}
\end{subequations}

\vskip 80pt
\begin{center}
\begin{picture}(35,25)
\put(-190,0){\resizebox{!}{3.3cm}{\includegraphics{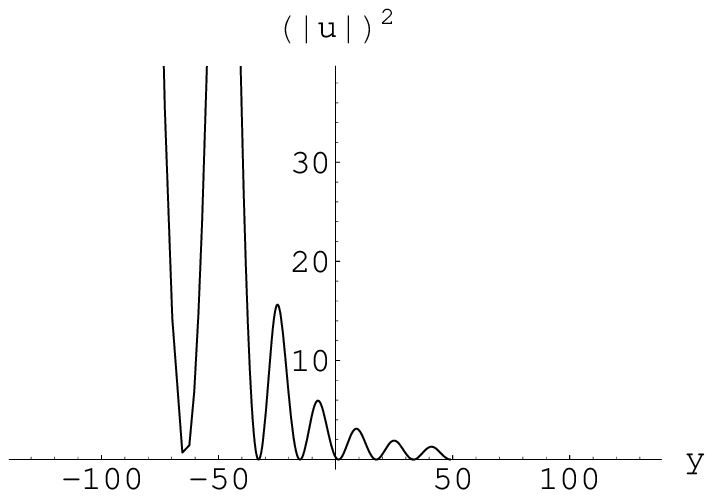}}}
\put(-40,0){\resizebox{!}{3.3cm}{\includegraphics{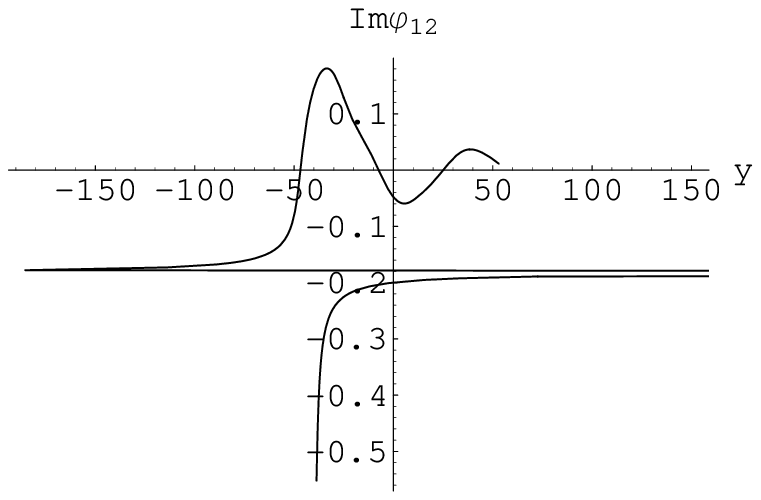}}}
\put(120,0){\resizebox{!}{3.3cm}{\includegraphics{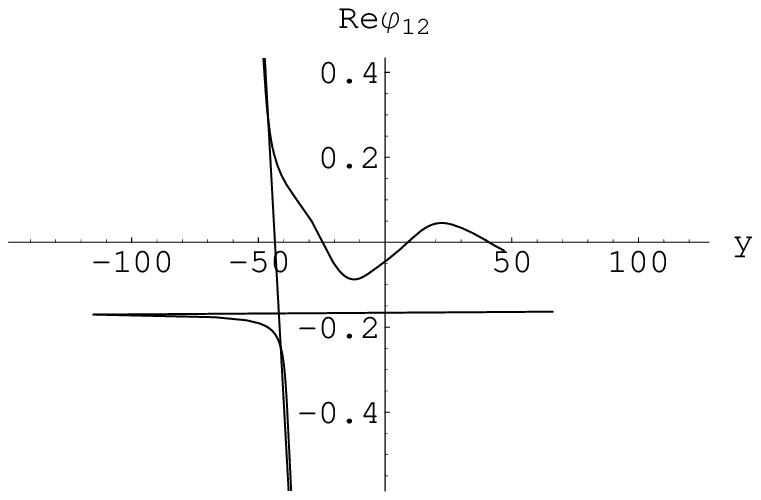}}}
\end{picture}
\end{center}
\begin{center}
\begin{minipage}{ 14cm}{\footnotesize
~~~~~~~~~~~~~~~~~~~~~~(a)~~~~~~~~~~~~~~~~~~~~~~~~~~~~~~~~~~(b)~~~~~~~
~~~~~~~~~~~~~~~~~~~~~~~~~~~~~~~~~~~~~~~~~(c)\\
{\bf Figure 8.}  The shapes  for $|u|^{2}$, the real part and
imaginary part of $\varphi_{12}$ with $\lambda_{1}=0.1,~~e(s)=2s,$
$s=2$, respectively. }
\end{minipage}
\end{center}

The shapes for the one positon solutions are given in Fig.8. By
using the N-negaton solutions and N-positon solutions of IDSGE and
the inverse reciprocal transformation (\ref{s22}), we can find the
N-negaton solutions and N-positon solutions for IDSPE. In the
reduced case we can find the new N-negaton solutions and N-positon
solutions for short pulse equation by taking all $e_j(s)$ to be
constants.

\section{Conclusion}
We first derive the integrable deformed short pulse hierarchy and
their zero curvature representation. Then we concentrate on the
solution of integrable deformed short pulse equation (IDSPE). By
proposing a generalized reciprocal transformation, we find a new
integrable deformed sine-Gordon equation (IDSGE) and its zero
curvature representation. The bilinear equation  and the Wronskian
determinant solutions for the IDSGE are  given. Furthermore, based
on the inverse reciprocal transformation and the solutions of the
IDSGE, the N-loop soliton solutions, N-negaton and N-positon
solutions of the IDSPH are worked out. In the reduced case the new
N-negaton solutions and new N-positon solutions for short pulse
equation are obtained.

\section*{Acknowledgement}
The first author is very grateful to Prof. D. Y. Chen for his help.
This work is supported by National Basic Research Program of China
(973 Program) (2007CB814800), National Natural Science Foundation of
China (10901090,10801083) and  Chinese Universities Scientific Fund
(2009JS42,2009-2-05)

\end{document}